\documentclass[twoside,leqno,twocolumn]{article}

\usepackage{ltexpprt}
\usepackage{booktabs} 
\usepackage{amsmath, amssymb, bbm, setspace, mathrsfs, color, listings, graphicx, multicol, dcolumn, dsfont, subcaption}
\usepackage[font=footnotesize,labelfont=bf,skip=5pt]{caption}
\usepackage{float,subcaption}
\usepackage[linesnumbered,ruled,algo2e]{algorithm2e}
\usepackage{cite}
\usepackage{url}
\allowdisplaybreaks

\newcommand{\diag}[1]{\text{diag}\left( #1 \right)}
\newcommand{\invgamma}[1]{\Gamma^{-1}\left( #1 \right)}

\DeclareMathOperator*{\rank}{rank}
\DeclareMathOperator*{\N}{N}
\DeclareMathOperator*{\Unif}{Unif}
\DeclareMathOperator*{\median}{median}

\begin{document}

\title{\large ABACUS: Unsupervised Multivariate Change Detection via Bayesian Source Separation}
\author{Wenyu Zhang\thanks{Cornell University}\\wz258@cornell.edu\\
\and
Daniel Gilbert\footnotemark[1]\\deg257@cornell.edu\\
\and 
David S. Matteson\footnotemark[1]\\matteson@cornell.edu}
\date{}

\maketitle

\begin{abstract}
Change detection involves segmenting sequential data such that observations in the same segment share some desired properties. 
Multivariate change detection continues to be a challenging problem due to the variety of ways change points can be correlated across channels and the potentially poor signal-to-noise ratio on individual channels. 
In this paper, we are interested in locating additive outliers (AO) and level shifts (LS) in the unsupervised setting. We propose ABACUS, {\it Automatic BAyesian Changepoints Under Sparsity}, a Bayesian source separation technique to recover latent signals while also detecting changes in model parameters. Multi-level sparsity achieves both dimension reduction and modeling of signal changes. 
We show ABACUS has competitive or superior performance in simulation studies against state-of-the-art change detection methods and established latent variable models. We also illustrate ABACUS on two real application, modeling genomic profiles and analyzing household electricity consumption.


\end{abstract}

\section{Introduction}

Change detection segments sequential data such that observations in each segment share the same characteristics. We can view it as a specific form of clustering where sequential data points tend to cluster together. Two common sequential orderings are time and physical location. 
Offline change detection segments the data retrospectively and is useful for uncovering events and systematic behaviors in data analysis tasks. It is applied in a variety of fields including energy consumption \cite{Harlé16}, genomics \cite{olshen04} and finance \cite{erdman07}. Furthermore, in the potential presence of change points, utilizing change detection prior to data modeling can help prevent building inappropriate models under the assumption of data homogeneity, and consequently supports improved prediction and statistical inference. 

\begin{figure}
\includegraphics[width=0.9\linewidth]{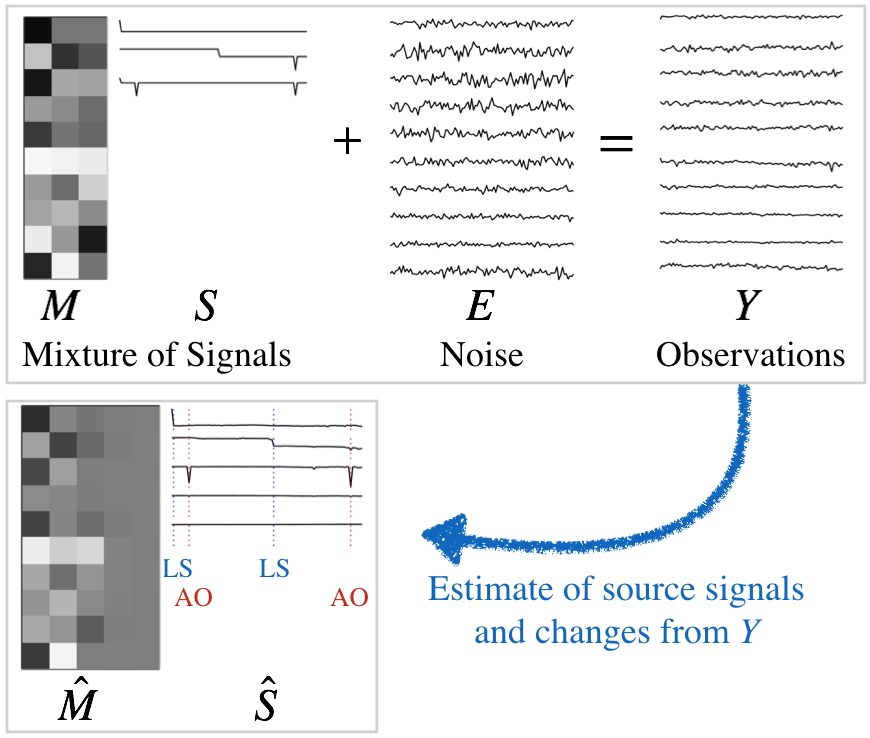}
\caption{Given observations generated by the linear mixing of signals contaminated by noise, ABACUS estimates the source signals and detect additive outliers (AO, red) and level shifts (LS, blue). In $M$, darker and lighter cells represent negative and positive values respectively, and medium gray cells represent zero.}
\label{fig:overview}
\end{figure}

In this paper, we study offline multiple change detection in multivariate data, specifically where the data exhibit mean changes that can occur simultaneously in several channels. The direction and magnitude of change can be different across channels. Here, we refer to mean changes lasting a single time unit with an immediate return as additive outliers (AO), and mean changes with duration two or greater as level shifts (LS). We assume that the multivariate data are generated by low-dimensional latent source signals through linear mixing according to the model $Y=MS+E$, shown in Figure \ref{fig:overview}, similar to the general linear setting used in the blind source separation literature \cite{hyvarinen01,matteson17}. Notation-wise, $M$ is the mixing matrix, and $Y$, $S$ and $E$ are the observations, source signals and noise, respectively. Observed mean changes manifest from the latent space, and we detect changes by estimating these latent source signals, which possess `semantic' meaning of the underlying states and are free of noise.

Multivariate data are readily observed in many applications in today's world, and mean changes are of particular interest since the mean is often a salient aspect of the system state. Multivariate data can be observations from multiple channels monitoring a single system, or a collection of univariate data streams from multiple related systems. Examples of the first scenario include household power consumption measured with sub-meters \cite{Harlé16}, and wine quality based on physicochemical test variables \cite{alippi16}. Examples of the second scenario include array comparative genomic hybridization measurements from several patients with the same medical condition \cite{matteson14}. In these and other examples, change points in multivariate data sometimes occur simultaneously in multiple channels because the signals may be driven by the same underlying processes. It is of interest to identify these shared change points to further analyze the relationship between channels. Running univariate change detection on each channel does not encourage identification of such shared changes.

Finding changes in multidimensional data is known to be a difficult problem. If the magnitude of change as measured by symmetric Kullback-Leibler divergence is kept constant, detectability of the change worsens when the data dimension $P$ increases. This can hinder detection even at dimensions as low as $P=10$ \cite{alippi16}. Another issue arises when the data dimensions $P$ exceeds the sample size $N$. If one wishes to use hypothesis testing to test for homogeneity, naive calculations of familiar test statistics such as the Hotelling's t-squared statistic are prohibitive. Several approaches tackle multivariate data by incorporating a dimensionality reduction step \cite{xie15,wang18}, but these either project the data onto a single dimension or require the user to select the reduced dimensionality.

Our main contribution is to successfully integrate sparse Bayesian blind source separation with a change detection framework. No previous work on latent variable modeling explicitly considered source signals with unconstrained mean changes. Bayesian variations of principle component analysis (PCA) are capable of automatic dimensionality selection \cite{bishop98,zhao14}, and shrinkage priors also achieve desirable properties in trend filtering \cite{kowal18}. In our Bayesian latent model, we use horseshoe priors to recover the lower-dimensional source signals and to simultaneously model the change points. The two tasks complement each other since the source signals exhibit changes. We propose ABACUS, {\it Automatic BAyesian Changepoint Under Sparsity}, an automatic procedure that simultaneously detects additive outliers and level shifts via estimating components from the source separation problem. Figure \ref{fig:overview} gives an example where ABACUS recovers the true latent change space of size three by estimating values in the appropriate dimensions of $M$ and $S$ to zero, and ABACUS also locates relevant change points. We show through simulations and real data applications that ABACUS achieves better performance in both change detection and source recovery.

\section{Related Works}

Authors of \cite{bleakley11} formulated multivariate change detection as a group fused Lasso, and showed empirically that detection probability approaches one with increasing $P$ when noise is small. Variants of binary segmentation produce approximately optimal segmentations by iteratively detecting single change points \cite{olshen04, matteson14}. Dynamic programming with a suitable multivariate goodness-of-fit metric can recursively the data \cite{zhang17}. The above methods directly segment the observations and some assume independence across channels \cite{wang18, fryzlewicz14}. We recover the latent change space with prior belief that only the latent signals are independent given model parameters.

Some works use a two-step procedure with data compression onto a low dimension $K \ll P$ followed by change detection. Projection onto a single dimension enables univariate change detection \cite{fryzlewicz14}. For $K>1$, \cite{popescu10} applies univariate change detection on each latent signal after Independent Component Analysis (ICA). Random projection where the projection is either fixed or varied across time has been paired with hypothesis testing \cite{xie15}. Using compressive measurements, where the projection matrix is a random projection or drawn from a Gaussian ensemble, \cite{atia15} derives the number of observations required for a target detection delay. 
For the above methods, the user needs to specify the compression ratio through $K$. Our proposed method ABACUS is more robust to the specification of $K$ due to automatic dimensionality selection by our sparsity assumptions. In contrast to the latent variable model that we employ, these methods also ignore estimating the mixing matrix.

Bayesian approaches in change detection typically rely on using indicator variables to denote the presence of change points. The BCP method \cite{erdman07, barry93} assumes that observations in each segment are independent and identically distributed as Gaussian, and updates posterior segment means conditional on the segmentation at each iteration of an MCMC scheme. A uniform prior $\text{U}(0, q)$ is put on the change point probabilities, and the user tunes the chances of discovering shorter or longer segments through $q$. In \cite{Harlé16}, given the segmentation informed by the indicator variables, a Wilcoxon rank sum test is performed at each index of the data and the resulting p-values are modeled as a Beta-Uniform mixture. The data likelihood is written as a composite marginal likelihood of the p-values. The formulation makes no assumption on the distributional form of the data. 

ABACUS similarly utilizes the sparsity of changes by applying horseshoe priors, modeling the presence and absence of changes, but also the change directions and magnitudes. We utilize the horseshoe prior as it is known for robustness and superior shrinkage properties \cite{carvalho09}. Empirically, differences in neighboring non-change location means are effectively shrunk to zero.

\section{Problem Formulation}
\label{sec: problem}

We observe $Y\in \mathbbm{R}^{P\times N}$, a $P$-dimensional data stream of length $N$. Each column take the form $Y_{\cdot n} = MS_{\cdot n} + E_{\cdot n}$, where $M\in \mathbbm{R}^{P\times r}$ is the mixing matrix, $S_{\cdot n}$ is the $r$-dimensional source signal, and $E_{\cdot n}$ is the $P$-dimensional noise vector, at index $n$. This is the general formulation of the cocktail party problem with $P$ microphones and $r$ conversations observed for $N$ time points. Here, $Y$ is not necessarily a time series, but data which are indexed sequentially. $S$ is assumed to have full row rank.

We assume that the source signals are piecewise-constant. Each segment can be of any length, and adjacent segments have different means. Latent variables are driven by the same underlying system state, and hence may share change locations, but change directions and magnitudes are not necessarily the same. We assume that the linearly-mixed signals are corrupted by independent Gaussian noise, but noise variances are not necessarily the same across channels. In the cocktail party analogy, this means that each microphone is subject to a different amount of noise due to the environment and microphone quality. The Gaussian assumption is standard in parametric change detection models \cite{xie15, olshen04, barry93}.

We aim to decompose $Y$ into its components without further information. Although the decomposition solution is not unique, \cite{gao13} reports that sparsity formulations in their Bayesian latent variable model helped to stabilize fitting. We similarly apply multiple levels of sparsity in our model, as described in the next section.

\section{Proposed Method: ABACUS}

We introduce our Bayesian data model and estimation method, as well as our change detection approach which makes use of MCMC posterior samples.

\subsection{A Bayesian Latent Variable Model}
\label{sec: bayesian model}

We decompose source signals further into components consisting of either additive outliers (AO) or level shifts (LS). Additive outliers are abrupt mean changes lasting for only one index, while level shifts persist for two or more indices. This decomposition allows us to naturally distinguish between the two types of changes, such that they can be studied separately, e.g., a user may remove additive outliers and retain level shifts for analysis. Let $K$ be a user-specified upper bound for $\rank(S)=r$ such that $r\leq K < P$. Then our modified formulation is
\begin{align*}
Y_{\cdot n} &= MS_{\cdot n} + E_{\cdot n} \\
S_{\cdot n} &= S^{(0)}_{\cdot n} + S^{(1)}_{\cdot n} \\
S^{(0)}_{\cdot n} &= V^{(0)}_{\cdot n} \text{ and }
\triangle S^{(1)}_{\cdot n} = V^{(1)}_{\cdot n}
\end{align*}
where $M$ is the $P\times K$ mixing matrix, $S$ is the $K\times N$ source signal matrix, $E$ is the $P\times N$ error matrix, $S^{(0)}$ and $S^{(1)}$ are the $K\times N$ component matrices of $S$, $V^{(0)}$ and $V^{(1)}$ are $K\times N$ `sparse' matrices, and $\triangle$ is the differencing operator. The diagonal covariance matrix of $E_{\cdot n}$ is denoted by $\Psi = \diag{\psi}$, so $E_{\cdot n} \sim \N(0, \ \Psi)$.

We place sparse group priors on the columns of $M$ and rows of $V^{(0)}$ and $V^{(1)}$ for dimensionality reduction of the latent space. Furthermore, we place sparse group priors on the columns of $V^{(0)}$ and $V^{(1)}$ to select a subset of indices as change locations. We also use elementwise sparsity on $V^{(0)}$ and $V^{(1)}$ to allow sparse changes for each latent variable.

We choose to use horseshoe priors because the horseshoe-shaped shrinkage profile discovers null values without diminishing strong signals. \cite{carvalho09}. We extend the global-local shrinkage hierarchy to impose sparsity in the model at the element and group level.

For $1 \leq i \leq P$ and $1 \leq h \leq K$ and $1 \leq n \leq N$ and $d \in \{0, 1\}$, we set priors as
\begin{align*}
M_{\cdot h}|\lambda_h^{(0)}, \lambda_h^{(1)}, \ \tau^{(0)}, \tau^{(1)}, \Psi &\sim \N\left(0, \lambda^{(0)}_h \lambda^{(1)}_h \tau^{(0)} \tau^{(1)} \Psi\right) \\
V^{(d)}_{hn}|\phi^{(d)}_n, \lambda^{(d)}_h, \gamma^{(d)}_{hn}, \ \tau^{(d)} &\sim \N\left(0, \phi^{(d)}_n \lambda^{(d)}_h \gamma^{(d)}_{hn} \tau^{(d)}\right) \\
\psi_i &\sim \invgamma{1, \ 1} \\
\tau^{(d)}|\xi^{(d)} &\sim \invgamma{\frac{1}{2}, \ \frac{1}{\xi^{(d)}}} \\
\lambda_h^{(d)}|\eta^{(d)}_h &\sim \invgamma{\frac{1}{2}, \ \frac{1}{\eta^{(d)}_h}} \\
\phi^{(d)}_n|\omega^{(d)}_n &\sim \invgamma{\frac{1}{2}, \ \frac{1}{\omega^{(d)}_t}} \\
\gamma^{(d)}_{hn}|\zeta^{(d)}_n &\sim \invgamma{\frac{1}{2}, \ \frac{1}{\zeta^{(d)}_{hn}}} \\
\xi^{(d)}, \eta^{(d)}_h, \omega^{(d)}_n, \zeta^{(d)}_{hn} &\sim \invgamma{\frac{1}{2}, \ 1}
\end{align*}
where $N()$ denotes the Gaussian distribution and $\Gamma^{-1}()$ denotes the Inverse Gamma distribution. Marginally, the shrinkage parameters $\tau^{(d)}, \lambda^{(d)}_h$, $\phi^{(d)}_n$ and $\gamma^{(d)}_{hn}$ are half-Cauchy, as in the horseshoe setup. Given the shrinkage parameters, we impose the prior belief that the source signals are independent, but the posterior is not necessarily so.

Let $D^{(1)}$ be the matrix representation of $\triangle$ such that $S^{(1)} \left[D^{(1)}\right]^T = V^{(1)}$, and let $D^{(0)} = I$ such that $S^{(0)} = V^{(0)}$. Now, we define the expression $F = SS^T+\diag{\tau^{(0)} \tau^{(1)} \lambda^{(0)} \lambda^{(1)}}^{-1}$, which appears below. 

For $1 \leq i \leq P$, $1 \leq n \leq N$, and $d \in \{0, 1\}$, we derive the full conditionals for the posterior distribution of the main model components below. Distributions of all additional parameters are provided in the Supplementary Materials. First,
\begin{align*}
M_{i\cdot}|\cdot &\sim \N\left( F^{-1}SY_{i\cdot}, \ \psi_i F^{-1} \right) \\
\psi_i|\cdot &\sim \invgamma{1+\frac{N}{2}, 
	\ 1 + \frac{1}{2}(Y_{i\cdot} - M_{i\cdot}S)^T (Y_{i\cdot} - M_{i\cdot}S)}
\end{align*}
and for $V^{(d)}_{\cdot n}$, the full conditional distribution is
\begin{equation*}
	\N\left( \left[ B^{(n)} \right]^{-1} M^T \Psi^{-1} C^{(n)} \left[D^{(d)}\right]^{-1}_{\cdot n}, 
	\ \left[ B^{(n)} \right]^{-1} \right)
\end{equation*}

where
\begin{align*}
B^{(n)} &= M^T \Psi^{-1} M \left( \left[D^{(d)}\right]^{-T}_{n\cdot} \left[D^{(d)}\right]^{-1}_{\cdot n} \right) + \\
    &\hspace{3cm} \diag{\phi^{(d)}_n \lambda^{(d)} \gamma^{(d)}_{\cdot n} \tau^{(d)}}^{-1} \\
C^{(n)} &= Y - MS + MV^{(d)}_{\cdot n}\left[D^{(d)}\right]^{-T}_{n\cdot}.
\end{align*}

We use Gibbs sampling to approximate the posterior. The procedure is easily parallelized. Furthermore, the number of model components and parameters depend on $K$ and correctly setting a small $K$ can significantly reduce computational time.

In our modified $Y = MS + E$ model, multiple levels of sparsity regulate the transformations each solution pair $M$ and $S$ can take to reach a different solution pair, but we cannot identify the sign and scaling of $M$ and $S$. To recover the components and parameters empirically, we use the median of the posterior samples to provide robustness against possible movements of the sampling path between different solutions. 

\subsection{Change Detection}
\label{sec: detection}

In our data model, $V^{(0)}$ and $V^{(1)}$ contain the changes for each latent variable at each index. The matrices are sparse since only entries which correspond to changes are nonzero. Let $f^{(d)}_n$ be the element with the largest magnitude in $V^{(d)}_{\cdot n}$. At any index $n$, $f^{(d)}_n$ is nonzero if and only if there is a change of type $d$ in at least one latent variable. Finding all such indices is equivalent to finding the change locations. We use the median defined
\begin{equation*}
    \widehat{g}^{(d)}_n = \median\left( \widehat{f}^{(d)}_n \right)
\end{equation*}
for robustness with empirical samples. 

Since we impose horseshoe priors on $V^{(d)}$, the entries are shrunk to approximately zero but not exactly zero. To identify the approximately zero values in the estimated $\widehat{g}^{(d)}$, we apply kernel density estimation on $|\widehat{g}^{(d)}|$ with a rectangular kernel and set the cutoff to be at the first minimum in the density function such that the minimum value is below threshold $\delta$. The threshold ensures that the approximately zero and non-zero values are sufficiently different. We set $\delta = 10^{-10}$ for all our experiments.

\section{Implementation}

We fit the full Bayesian latent variable model in Section \ref{sec: bayesian model} by first fitting a partial model. The partial model differs only in that it does not include $S^{(0)}$ or $V^{(0)}$ and their associated parameters, and hence we drop the superscripts when referring to its components and parameters. Changepoints $cpt$ detected by the partial model are a mix of additive outliers (AO) and level shifts (LS), with the former being detected as two consecutive mean changes of opposite signs in $\widehat{g}$. We distinguish between the two types of changes according to this observation with Algorithm \ref{alg: separate}, and produce additive outliers $cpt0$ and level shifts $cpt1$. We decompose the estimated components and parameters from the partial model according to $cpt0$ and $cpt1$, and pass them to the full model as initialization. For example, $V^{(0)}$ is initialized with values from $\widehat{V}$ at $cpt0$, and $V^{(1)}$ is initialized with values from $\widehat{V}$ at $cpt1$.

\begin{algorithm2e}
 \KwData{Estimated $\widehat{g}$, ordered change points $cpt$}
 \KwResult{Additive outliers $cpt0$, level shifts $cpt1$}
 $cpt0 = cpt1 = \{\}$\;
 $i = 1$\;
 \While{not at end of $cpt$}{
    condition 1: $cpt[i+1] - cpt[i] = 1$\;
    condition 2: $\widehat{g}$ corresponding to $cpt[i]$ and 
        $cpt[i+1]$ are of opposite signs\;
    \eIf{condition 1 and 2 are True}{
        add $cpt[i]$ to $cpt0$\;
        $i = i+2$\;
        }{
        add $cpt[i]$ to $cpt1$\;
        $i = i+1$\;
        }
 }
 \caption{Separating AO and LS changes}
 \label{alg: separate}
\end{algorithm2e}

\begin{figure*}
\includegraphics[width=\linewidth]{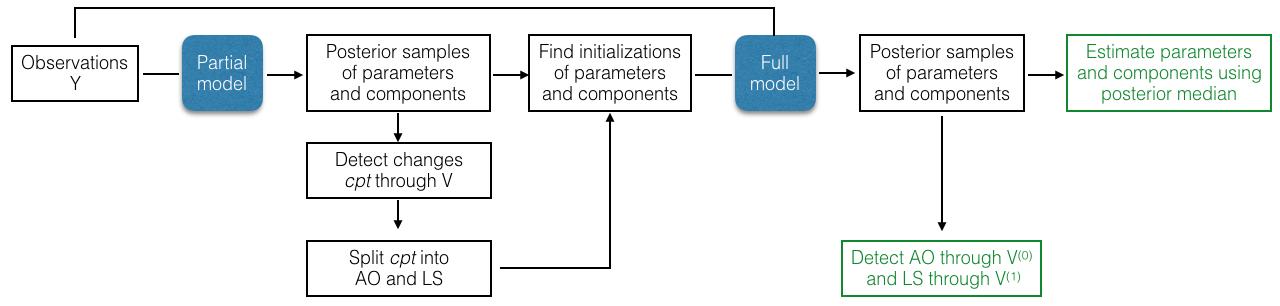}
\caption{Implementation procedure. From observations Y, a partial model is first fit and its estimations initialize the full Bayesian model. Final estimates of source signals and change points are obtained from the median of MCMC samples.}
\label{fig:alg}
\end{figure*}

The partial model is smaller and hence can quickly estimate components and parameters for initialization.
This step stabilizes fitting the full model and helps to achieve better distinction between the two types of changes. The entire procedure is shown in Figure \ref{fig:alg}. The final two boxes in green indicate the final outputs for change detection and source recovery.

\section{Simulation Study}
\label{sec: simulation}

We conduct several experiments according to the model $Y = MS + E$ described in Section \ref{sec: problem}. We fix the latent space dimensionality $r = 3$, and vary $N$ and $P$. Some methods require a user-specified $K$ as an estimate for $r$, and we test their robustness to the selection of $K$. Entries of $M$ are drawn independently from $\Unif(-1,1)$, and each noise variance as $\psi_i \sim \Unif(0.1, 5)$. Given the number of additive outliers and level shifts, change locations are sampled uniformly at random from $\{2, 4, 6, \dots, N-1\}$. This ensures that level shifts are at least of length two and that we do not unintentionally construct level shifts through consecutive additive outliers. To construct sparse changes, at each change location, the number of latent signals experiencing change is selected uniformly at random. Change magnitudes are drawn from $\Unif(1,5)$ with the sign being equally likely to be positive or negative.

We compare ABACUS against state-of-the-art change detection techniques and popular latent variable models which are marked by $\times$ and $\circ$, respectively, in plots in this section. We use default parameters in software packages unless otherwise specified. To find additive outliers, we set the minimum segment length parameter to one where possible in competing change detection implementations. The detected changes are categorized into additive outliers and level shifts using Algorithm \ref{alg: separate} without Condition 2, except for TSO mentioned below which automatically outputs different types of changes. For all MCMC procedures, number of iterations is $3000$ and burn-in is $500$. Each simulation is run $100$ times, and we report the average performance according to the evaluation metrics in Section \ref{sec: criteria}.

Amongst competing multivariate change detection methods, GFLseg \cite{bleakley11} finds candidate mean changes by group fused Lasso followed by selection via dynamic programming. E-divisive \cite{matteson14} uses binary segmentation to iteratively locate each single change point through measuring between-segment distance by the energy statistic. We specify its moment index parameter $\alpha = 2$ to find level shifts, and $min.size = 2$ the smallest segment length allowed, which implies E-divisive is unable to find additive outliers. BCP \cite{erdman07} is a Bayesian method which models the presence of mean change at each location through an indicator variable and uses MCMC sampling to infer the posterior probability of change. BCP outputs a set of change points corresponding to each posterior sample, hence for evaluation we compute the average metric across all these sets. We also combine BPCA \cite{bishop98} and BCP to obtain a two-step Bayesian approach to first compress and then detect. 

Inspect \cite{wang18} transforms observations into a univariate series through cumulative sum transformation before applying wild binary segmentation. We also test three univariate methods by first applying PCA to the observations. PELT \cite{killick12} is a popular parametric approach that uses dynamic programming to efficiently find the segmentation that minimizes the negative log-likelihood plus a penalty. We refer to the non-parametric version as np-PELT, which uses the empirical distribution instead \cite{haynes16}. A third method, TSO, jointly estimates ARIMA model parameters and change effects due to additive outliers and level shifts \cite{lacalle17}.

To fit the latent variable model, we tested against well-established methods including Independent Component Analysis (ICA), Factor Analysis (FA) and Bayesian Principal Component Analysis (BPCA). Note that ICA and FA do not impose sparsity assumptions, whereas BPCA imposes sparsity on the columns of $M$. For ICA, we use the FastICA implementation which measures non-Gaussianity using negentropy \cite{hyvarinen00}. For FA, we use the factanal function in R \cite{R18} which automatically checks for identifiability given $K$ and does not fit a model if $K$ is too large to fit a unique model.

\subsection{Evaluation Criteria}
\label{sec: criteria}

We evaluate the detection of additive outliers and level shifts separately since some competing methods \cite{matteson14} detect one but not the other. We report precision and recall, and treat an estimate as accurate if it is within $w$ of a true change location. We set $w = 1$ for the small sample experiment in Section \ref{sec: sim1}, and $w = 3$ for the larger sample experiments in Section \ref{sec: sim2} and \ref{sec: sim3}.

We evaluate the quality of model recovery through components $M$ and $S$, and noise variance parameter $\psi$. Given true mixing matrix $M$ and estimate $\widehat{M}$, we center and scale each row of the matrices and measure their dissimilarity using the squared trace metric in \cite{gao13}, 
\begin{equation*}
    \epsilon_M = \frac{1}{P^2}Tr\left( MM^T - \widehat{M}\widehat{M}^T \right).
\end{equation*}
The metric $\epsilon_M$ is invariant to orthogonal rotation and allows cases where either $MM^T$ or $\widehat{M}\widehat{M}^T$ is singular. Next, given true source signals $S$ and estimate $\widehat{S}$, we measure their dissimilarity using a variation of averaged squared Euclidean distance
\begin{equation*}
    \epsilon_S = \frac{1}{r}\sum_{i=1}^r \left( 1-|\rho_i| \right)
\end{equation*}
where $\rho_i$ is the Pearson correlation coefficient between $S_{i\cdot}$ and some $\widehat{S}_{j\cdot}$, and each pair is found greedily by descending magnitude of correlation. This measure is invariant to sign and label switching. Finally, given true noise variance $\psi$ and estimate $\widehat{\psi}$, the difference is measured by their scaled squared norm
\begin{equation*}
    \epsilon_E = \frac{1}{P}\|\psi - \widehat{\psi}\|_2^2.
\end{equation*}

\subsection{Simulation 1: Variations in $P$}
\label{sec: sim1}

We test the case of small sample size $N = 100$ and varying $P \in \{10, 30, 60, 90, 110\}$. Each sample has two additive outliers and two level shifts, and $K$ is set to $5$.

\begin{figure*}

\begin{subfigure}[b]{.5\linewidth}
\includegraphics[width=0.49\linewidth]{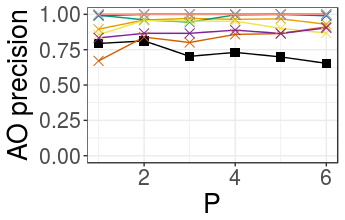}
\includegraphics[width=0.49\linewidth]{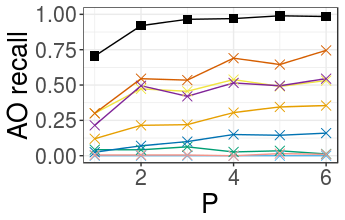}
\caption{Additive outliers}
\end{subfigure}
\begin{subfigure}[b]{.5\linewidth}
\includegraphics[width=0.49\linewidth]{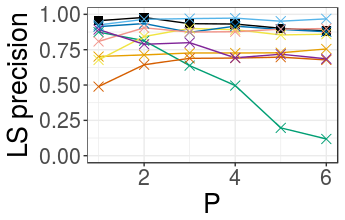}
\includegraphics[width=0.49\linewidth]{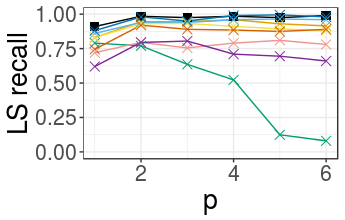}
\caption{Level shifts}
\end{subfigure}
\caption{Average errors in change detection as data dimensionality $P$ is varied; $N=100$ and $K=5$ are fixed.}
\label{fig:plotP_change}

\begin{subfigure}[b]{.24\linewidth}
\includegraphics[width=\linewidth]{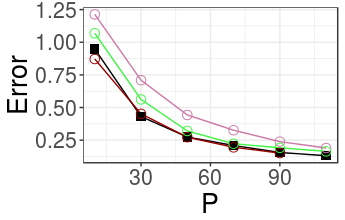}
\caption{Error $\epsilon_M$ for M}
\end{subfigure}
\begin{subfigure}[b]{.24\linewidth}
\includegraphics[width=\linewidth]{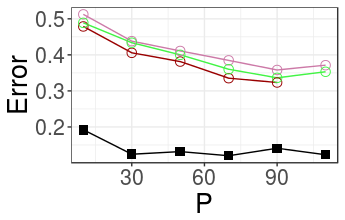}
\caption{Error $\epsilon_S$ for S}
\end{subfigure}
\begin{subfigure}[b]{.24\linewidth}
\includegraphics[width=\linewidth]{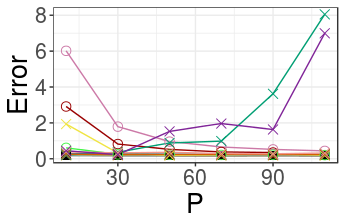}
\caption{Error $\epsilon_E$ for $\psi$}
\end{subfigure}
\begin{subfigure}[b]{.24\linewidth}
\includegraphics[width=\linewidth]{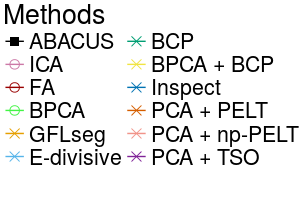}
\end{subfigure}
\caption{Average errors in model recovery as data dimensionality $P$ is varied; $N=100$ and $K=5$ are fixed. FA does not support computations for $P=110$ due to non-identifiability.}
\label{fig:plotP_model}

\begin{subfigure}[b]{.5\linewidth}
\includegraphics[width=0.49\linewidth]{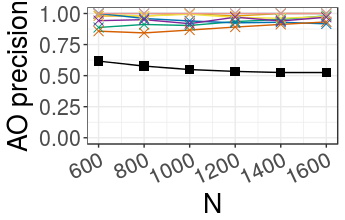}
\includegraphics[width=0.49\linewidth]{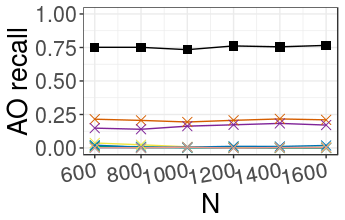}
\caption{Additive outliers}
\end{subfigure}
\begin{subfigure}[b]{.5\linewidth}
\includegraphics[width=0.49\linewidth]{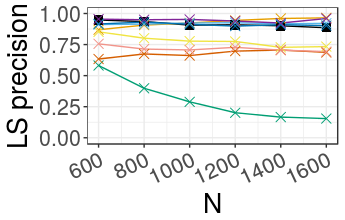}
\includegraphics[width=0.49\linewidth]{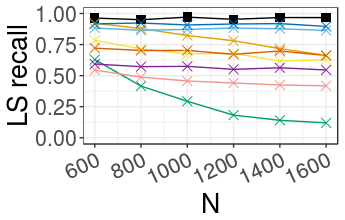}
\caption{Level shifts}
\end{subfigure}
\caption{Average errors in change detection as sample size $N$ is varied; $P=10$ and $K=5$ are fixed.}
\label{fig:plotN_change}

\begin{subfigure}[b]{.24\linewidth}
\includegraphics[width=\linewidth]{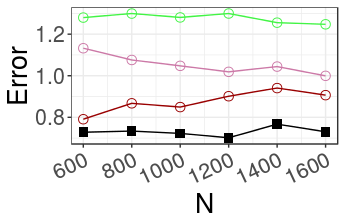}
\caption{Error $\epsilon_M$ for M}
\end{subfigure}
\begin{subfigure}[b]{.24\linewidth}
\includegraphics[width=\linewidth]{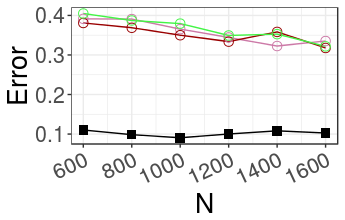}
\caption{Error $\epsilon_S$ for S}
\end{subfigure}
\begin{subfigure}[b]{.24\linewidth}
\includegraphics[width=\linewidth]{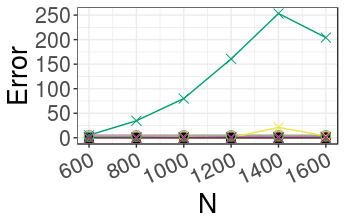}
\caption{Error $\epsilon_E$ for $\psi$}
\end{subfigure}
\begin{subfigure}[b]{.24\linewidth}
\includegraphics[width=\linewidth]{legend.png}
\end{subfigure}
\caption{Average errors in model recovery as sample size $N$ is varied; $P=10$ and $K=5$ are fixed.}
\label{fig:plotN_model}

\begin{subfigure}[b]{.5\linewidth}
\includegraphics[width=0.49\linewidth]{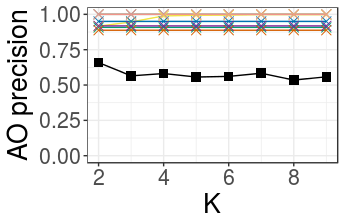}
\includegraphics[width=0.49\linewidth]{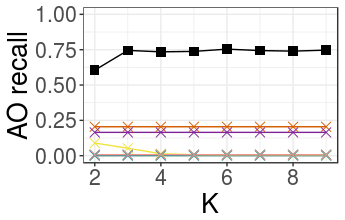}
\caption{Additive outliers}
\end{subfigure}
\begin{subfigure}[b]{.5\linewidth}
\includegraphics[width=0.49\linewidth]{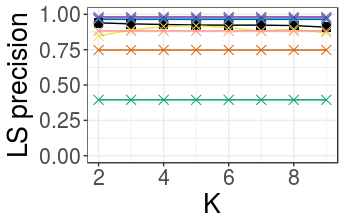}
\includegraphics[width=0.49\linewidth]{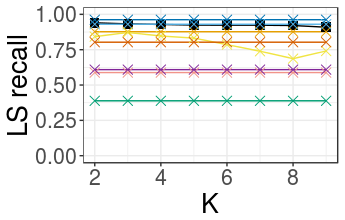}
\caption{Level shifts}
\end{subfigure}
\caption{Average errors in change detection as estimated latent space dimensionality $K$ is varied; fixed $N = 1000$ and $P=10$.}
\label{fig:plotK_change}

\begin{subfigure}[b]{.24\linewidth}
\includegraphics[width=\linewidth]{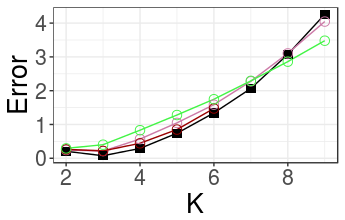}
\caption{Error $\epsilon_M$ for M}
\end{subfigure}
\begin{subfigure}[b]{.24\linewidth}
\includegraphics[width=\linewidth]{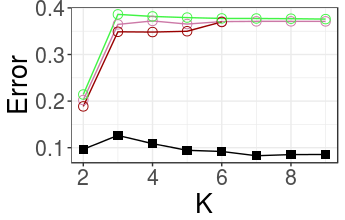}
\caption{Error $\epsilon_S$ for S}
\end{subfigure}
\begin{subfigure}[b]{.24\linewidth}
\includegraphics[width=\linewidth]{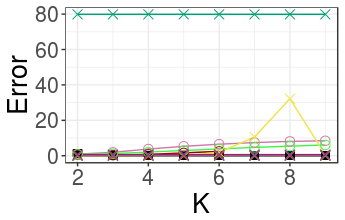}
\caption{Error $\epsilon_E$ for $\psi$}
\end{subfigure}
\begin{subfigure}[b]{.24\linewidth}
\includegraphics[width=\linewidth]{legend.png}
\end{subfigure}
\caption{Average errors in model recovery as latent space dimensionality parameter $K$ is varied; $N = 1000$ and $P=10$ are fixed. FA does not support computations for $K\geq 7$ due to non-identifiability.}
\label{fig:plotK_model}

\end{figure*}

As seen from Figure \ref{fig:plotP_change}, competing methods have high precision but low recall on additive outliers. As $P$ increases, ABACUS can locate most of the additive outliers, and is one of the best-performing methods for level shifts. Both precision and recall on level shifts decrease as $P$ increases for BCP, possibly because parameters such as the prior on change probabilities need to be adjusted. BPCA + BCP has more consistent performance, indicating the advantage of detecting changes on latent signals. In terms of model recovery, our method also gives the lowest errors for $M$, $S$ and $\psi$, see Figure \ref{fig:plotP_model}.

\subsection{Simulation 2: Variations in $N$}
\label{sec: sim2}

We fix $P=10$ and vary $N \in \{600, 800, 1000, 1200, 1400, 1600\}$. Each sample has $\frac{N}{100}$ additive outliers and $\frac{N}{100}$ level shifts, and $K$ is set to $5$. Performance of all methods is consistent across $N$, as shown in Figures \ref{fig:plotN_change} and \ref{fig:plotN_model}. BCP shows deteriorating performance in detecting level shifts just as it did in Section \ref{sec: sim1}, again possibly because model parameters need to be adjusted according to the sample size. Overall, ABACUS offers the best balance of precision and recall on additive outliers while all other competing change detection methods tend to miss them. ABACUS has the highest recall for level shifts, and almost always has the lowest errors for model recovery.

\subsection{Simulation 3: Variations in $K$}
\label{sec: sim3}

We fix $P=10$ and $N=1000$. Each sample has ten additive outliers and ten level shifts. We vary the user-specified estimate of the latent space dimensionality $K$ between $2$ and $9$. The true dimensionality $r$ is $3$. The horizontal lines in Figures \ref{fig:plotK_change} and \ref{fig:plotK_model} correspond to results of methods which do not have the parameter $K$. According to Figure \ref{fig:plotK_change}, the change detection results of ABACUS are consistent across $K$. From Figure \ref{fig:plotK_model}, ABACUS has much more consistent error $\epsilon_S$ in $S$ compared to competing latent variable models, whose $\epsilon_S$ increases sharply at $K \geq r$.

\section{Application to Real Data}

In both data applications below we set $K=5$ and also study the robustness of ABACUS to different $K$ values.

\subsection{aCGH Data}

Array-based comparative genomic hybridization (aCGH) is a technique for studying copy number alterations in event of diseases. We obtain the dataset from the R package ecp \cite{james14}, which has already removed sequences with more than $7\%$ missing values, and leaves 43 samples of different individuals with bladder tumor. Each sample has 2215 probes measuring the log2 ratio between the number of transcribed DNA copies from tumorous cells and from a healthy reference \cite{Harlé16}. A negative ratio indicates deletion, a positive ratio indicates amplification, and zero indicates an unaltered segment. We expect shared change locations for individuals with the same medical condition.

To reduce computations and ease visualization, we thin the samples by taking every $20^{th}$ value. We arrive at a dataset with $P=43$ and $N=111$. ABACUS takes approximately one minute to run on a standard desktop computer, and finds three additive outliers and seven level shifts. An additive outlier here indicates a shorter segment of genetic aberration compared to a level shift. A plot of all 43 samples with the estimated change points overlaid is in the Supplementary Materials.

\begin{figure}
\includegraphics[width=\linewidth]{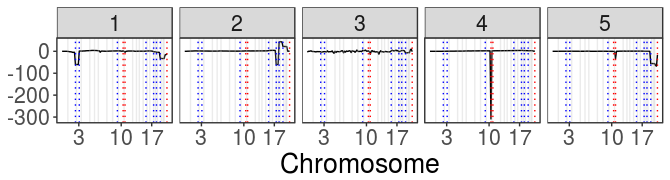}
\caption{aCGH: Latent source signals (1-5) recovered (black), and additive outliers (red) and level shifts (blue) detected. Gray lines indicate the boundaries between chromosome pairs.}
\label{fig:acgh_S}
\end{figure}

At least 99\% of the variance of our estimated latent signals can be explained by four principal components, while those found by ICA and FA require all five. As observed in Figure \ref{fig:acgh_S}, the third latent source signal recovered exhibits no evident changes. We map the four other signals to unique sets of genetic aberrations in different stages of bladder tumor in Table \ref{table:acgh}. For instance, patients with genetic aberrations on chromosome arms 2q, 3q and 20p/q simultaneously tend to be in tumor stage $pT_1$, hence the changes detected can be indicative of diseases for new patients.
The mapping is established based on a bladder tumor research article \cite{blaveri05} which lists the frequent genomic alterations by chromosome arm in tumor stages $pT_a$, $pT_1$ and $pT_{2-4}$. Each stage is determined pathologically depending on the size and location of the tumor. 

\begin{table}
  \begin{center}
      \begin{tabular}{ | l | p{4.8cm} | l |}
      \hline
      S & Chromosome arm with changes & Tumor stage \\ 
      \hline
      1 & 2q, 3q, 20p/q & $pT_1$ \\
      2 & 17p/q, 18p/q, 19p/q, 20p/q & $pT_1$ \\
      4 & 10q & $pT_a$, $pT_1$, $pT_{2-4}$ \\
      5 & 11p, 20p/q & $pT_{2-4}$ \\
      \hline
      \end{tabular}
  \end{center}
  \caption{aCGH: Genetic aberrations corresponding to changes detected on latent source signals. To read the table, 20p is the short arm of chromosome 20, and 20q is the long arm. Tumor stages range from $a$, $1$ to $4$ in order of severity.}
  \label{table:acgh}
\end{table}

\begin{figure}

\begin{subfigure}[b]{.49\linewidth}
\includegraphics[width=\linewidth]{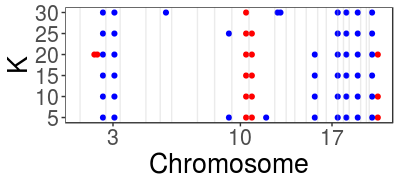}
\caption{Additive outliers (red) and level shifts (blue)}
\end{subfigure}
\begin{subfigure}[b]{.49\linewidth}
\includegraphics[width=\linewidth]{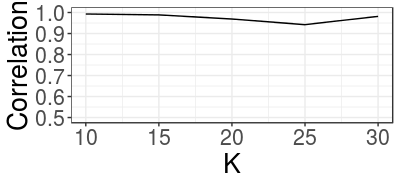}
\caption{Average correlation to latent signals at $K=5$}
\end{subfigure}
\caption{aCGH: Changes and latent source signals recovered by ABACUS are similar regardless of the specification of K.}

\label{fig:acgh_K}
\end{figure}

ABACUS performs consistently across different $K$.
Figure \ref{fig:acgh_K} shows that for $K\in\{10,15,20,25,30\}$, the change points and latent source signals recovered are very similar to those found with $K=5$.

\subsection{Electric Power Consumption Data}

This dataset contains per-minute measurements of electric power consumption in one household and is available on the UCI Machine Learning Repository \cite{dua17}. The data has seven dimensions including global active power (GAP), global reactive power (GRP), voltage (V), global intensity (GI), and three sub-meterings corresponding to the kitchen (S1), laundry room (S2) and heating system (S3). We expect shared change points since the seven dimensions are related arithmetically, and some electrical appliances tend to be used simultaneously. For instance, $\frac{1000}{60}\text{GAP}-\text{S1}-\text{S2}-\text{S3}$ is the power consumed by appliances outside of the kitchen, laundry room and heating system. We analyze a full day's worth of data, that is, the observation matrix has $P=7$ and $N=1440$. ABACUS takes approximately fifteen minutes to run on a standard desktop computer. 

The Supplementary Materials contain a plot of the standardized data with estimated changes overlaid. Although the data does not follow our model assumptions exactly since the amount of fluctuations or noise is more significant in the first half of the day, and there are minor trend changes in the second half of the day, ABACUS is robust and with post-processing it finds one additive outlier and sixteen level shifts. We post-process by dynamic programming to prune the initially estimated level shifts. This is similar to GFLseg \cite{bleakley11}, except that we apply the procedure on the latent source signals which are less contaminated by noise. 

\begin{figure}
\includegraphics[width=\linewidth]{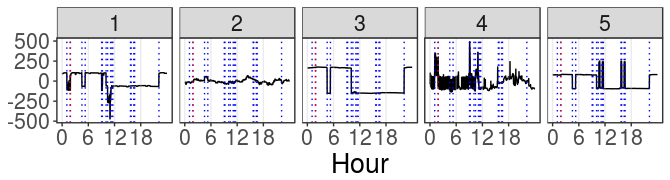}
\caption{Power: Latent source signals (1-5) recovered (black), and additive outliers (red) and level shifts (blue) detected.}
\label{fig:elec_S}
\end{figure}

\begin{figure}

\begin{subfigure}[b]{.49\linewidth}
\includegraphics[width=\linewidth]{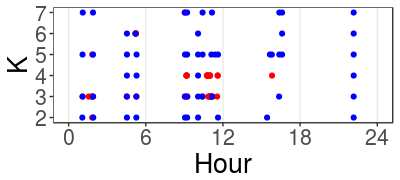}
\caption{Additive outliers (red) and level shifts (blue)}
\end{subfigure}
\begin{subfigure}[b]{.49\linewidth}
\includegraphics[width=\linewidth]{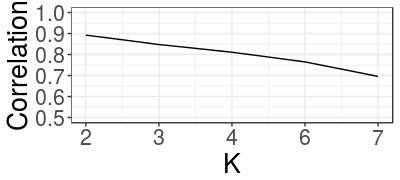}
\caption{Average correlation to latent signals at $K=5$}
\end{subfigure}
\caption{Power: Changes and latent source signals recovered by ABACUS are similar regardless of the specification of K.}

\label{fig:elec_K}
\end{figure}

The change points are indicative of the household's pattern of electricity usage, which concentrates in the first half of the day as illustrated in Figure \ref{fig:elec_S}. The fourth latent signal reflects the usage fluctuations and trends which differ across the two halves of the day as measured by GAP and GI.
ABACUS performs consistently across different specifications of $K$. Figure \ref{fig:elec_K} shows that for $K\in\{2,3,4,6,7\}$, the estimated change points and latent source signals recovered are similar to those found at $K=5$.

\begin{figure}
\centering
\includegraphics[width=0.8\linewidth]{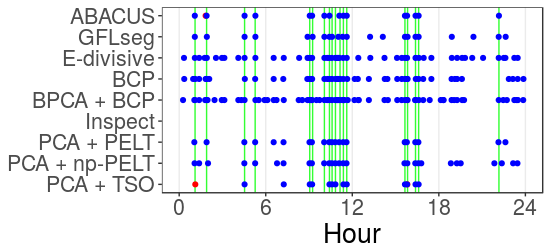}
\caption{Power: Additive outliers (red) and level shifts (blue) estimated vs ground truth level shifts (green).}
\label{fig:elec_cpt}
\end{figure}

\begin{figure}

\begin{subfigure}[b]{.49\linewidth}
\includegraphics[width=\linewidth]{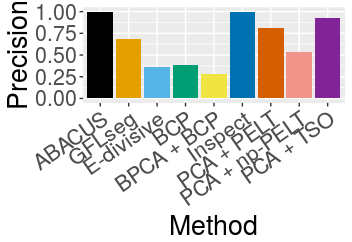}
\caption{Precision}
\end{subfigure}
\begin{subfigure}[b]{.49\linewidth}
\includegraphics[width=\linewidth]{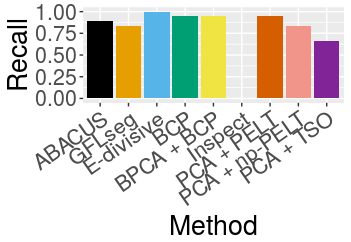}
\caption{Recall}
\end{subfigure}
\caption{Power: Performance in estimating level shifts.}

\label{fig:elec_perf}
\end{figure}

Since the sub-meterings S1, S2 and S3 demonstrate distinct level shifts when the respective appliances are utilized, we extract ground truths for level shifts by finding positions where these signals deviate from their base levels.
Compared to other change detection methods in Figure \ref{fig:elec_cpt} and \ref{fig:elec_perf}, ABACUS has the best overall performance with $\text{precision} = 1$ and $\text{recall} = 0.889$.

\section{Conclusion}

In this paper, we propose ABACUS, an automatic change detection procedure which makes use of Bayesian latent variable modeling. Due to the separation of additive outlier and level shift effects in the model, ABACUS naturally identifies these two types of changes separately, unlike many competing approaches.

In simulations, ABACUS shows competitive or superior performance in both change detection and model recovery. In two real data applications, ABACUS found relevant change points and source signals. It is robust to over-specification of $K$, an important property since the true value is rarely known to the user in practice.

\section{Acknowledgement}

NSF grant DMS-1455172, a Xerox PARC Faculty Research Award, and Cornell University Atkinson’s Center for a Sustainable Future AVF-2017 is gratefully acknowledged.

\bibliographystyle{siamplain}
\bibliography{biblio}

\appendix
\section{Supplementary Materials}

\subsection{Posterior Distributions}

Let $D^{(1)}$ be the matrix representation of $\triangle$ such that $S^{(1)} \left[D^{(1)}\right]^T = V^{(1)}$. Also $D^{(0)} = I$ such that $S^{(0)} = V^{(0)}$. We define the following expressions for the full conditionals of the posterior distributions:
\begin{align*}
F &= SS^T+\diag{\tau^{(0)} \tau^{(1)} \lambda^{(0)} \lambda^{(1)}}^{-1} \\
G^{(0)} &= \sum_{i=1}^p\sum_{h=1}^K \frac{M_{ih}^2}{2\lambda^{(0)}_h \lambda^{(1)}_h \tau^{(1)} \psi_i} +         \sum_{n=1}^N\sum_{h=1}^K \frac{\left[V^{(0)}_{hn}\right]^2}{2\phi^{(0)}_n \lambda^{(0)}_h \gamma^{(0)}_{hn}} \\
H^{(0)}_h &= \sum_{i=1}^p \frac{M_{ih}^2}{2\tau^{(0)} \tau^{(1)} \lambda^{(1)}_h \psi_i} + 
	\sum_{n=1}^N \frac{\left[V^{(0)}_{hn}\right]^2}{2\phi^{(0)}_n \gamma^{(0)}_{hn} \tau^{(0)}} \\
\end{align*}

For $1 \leq i \leq P$ and $1 \leq h \leq K$ and $1 \leq n \leq N$ and $d \in \{0, 1\}$, we derive the full conditionals for the posterior distributions below. We leave out $\tau^{(1)}$ and $\lambda_h^{(1)}$ since their full conditional distributions are similar in form to those of $\tau^{(0)}$ and $\lambda_h^{(0)}$ respectively.

\begin{align*}
M_{i\cdot}|\cdot &\sim \N\left( F^{-1}SY_{i\cdot}, \ \psi_i F^{-1} \right) \\
\psi_i|\cdot &\sim \invgamma{1+\frac{N}{2}, 
	\ 1 + \frac{1}{2}(Y_{i\cdot} - M_{i\cdot}S)^T (Y_{i\cdot} - M_{i\cdot}S)} \\
\tau^{(0)}|\cdot &\sim \invgamma{\frac{1+K(p+N)}{2}, \ \frac{1}{\xi^{(0)}} + G^{(0)}} \\
\xi^{(d)}|\cdot &\sim \invgamma{1, 1+\frac{1}{\tau^{(d)}}} \\
\lambda^{(0)}_h|\cdot &\sim \invgamma{\frac{1+p+N}{2}, \ \frac{1}{\eta^{(0)}_h} + H^{(0)}_h} \\
\eta^{(d)}_h|\cdot &\sim \invgamma{1, 1+\frac{1}{\lambda^{(d)}_h}} \\
\phi^{(d)}_n|\cdot &\sim \invgamma{\frac{1+K}{2}, 
	\ \frac{1}{\omega^{(d)}_n} + \sum_{h=1}^K \frac{\left[V^{(d)}_{hn}\right]^2}{2\lambda^{(d)}_h \gamma^{(d)}_{hn} \tau^{(d)}}} \\
\omega^{(d)}_n|\cdot &\sim \invgamma{1, 1+\frac{1}{\phi^{(d)}_n}} \\
\gamma^{(d)}_{hn}|\cdot &\sim \invgamma{1, 
	\ \frac{1}{\zeta^{(d)}_{hn}} +
	\frac{\left[V^{(d)}_{hn}\right]^2}{2\lambda^{(d)}_h \phi^{(d)}_n \tau^{(d)}}} \\
\zeta^{(d)}_{hn}|\cdot &\sim \invgamma{1, 1+\frac{1}{\gamma^{(d)}_{hn}}} \\
\end{align*}

For $V^{(d)}_{\cdot n}$, the full conditional distribution is
\begin{equation*}
	\N\left( \left[ B^{(n)} \right]^{-1} M^T \Psi^{-1} C^{(n)} \left[D^{(d)}\right]^{-1}_{\cdot n}, 
	\ \left[ B^{(n)} \right]^{-1} \right)
\end{equation*}

where
\begin{align*}
B^{(n)} &= M^T \Psi^{-1} M \left( \left[D^{(d)}\right]^{-T}_{n\cdot} \left[D^{(d)}\right]^{-1}_{\cdot n} \right) + \\
    &\hspace{3cm} \diag{\phi^{(d)}_n \lambda^{(d)} \gamma^{(d)}_{\cdot n} \tau^{(d)}}^{-1} \\
C^{(n)} &= Y - MS + MV^{(d)}_{\cdot n}\left[D^{(d)}\right]^{-T}_{n\cdot}
\end{align*}

\subsection{Additional Plots for aCGH Data}

\begin{figure*}
\includegraphics[width=\linewidth]{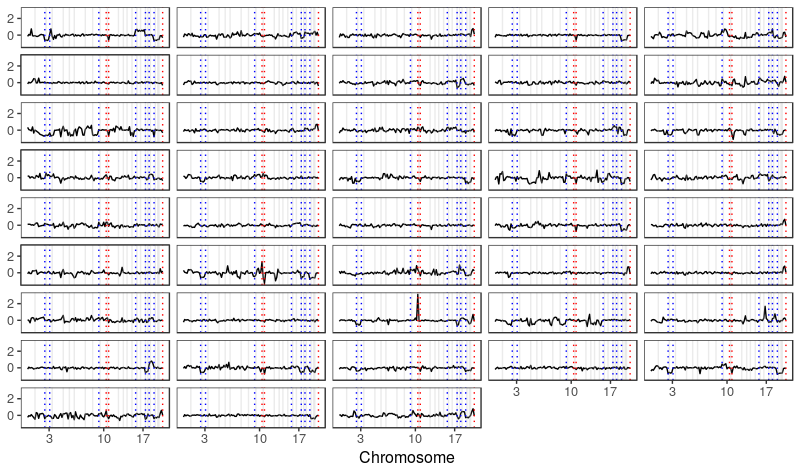}
\caption{aCGH: Additive outliers (red) and level shifts (blue) detected by ABACUS. Gray lines indicate the boundaries between chromosome pairs. Additive outliers correspond to shorter segments of genetic aberrations and level shifts correspond to longer segments.}
\label{fig:acgh}
\end{figure*}

Figure \ref{fig:acgh} plots all 43 samples with the estimated change points overlaid.

\begin{figure*}
\centering
\includegraphics[width=0.8\linewidth]{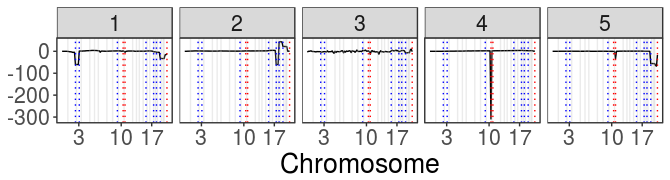}
\caption{aCGH: Latent source signals (1-5) recovered (black), and additive outliers (red) and level shifts (blue) detected. Gray lines indicate the boundaries between chromosome pairs.}
\label{fig:acgh_Snew}
\end{figure*}

For comparison, we include again the recovered latent source signals with the estimated change points overlaid in Figure \ref{fig:acgh_Snew}.

\subsection{Additional Plots for Electric Power Consumption Data}

\begin{figure*}
\includegraphics[width=\linewidth]{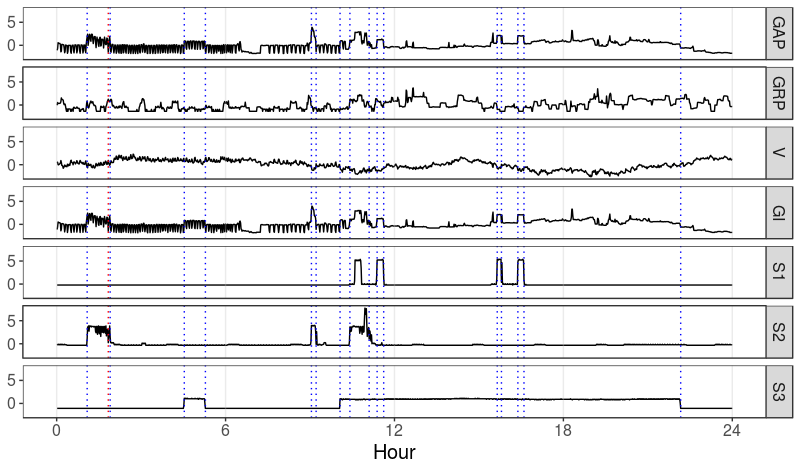}
\caption{Power: Additive outliers (red) and level shifts (blue) detected by ABACUS. The level shifts detected correspond well with appliance usages in sub-meterings S1, S2 and S3.}
\label{fig:elec}
\end{figure*}

Figure \ref{fig:elec} plots the standardized data from the electric power consumption dataset with estimated changes overlaid. ABACUS is run on the standardized dataset.

\begin{figure*}
\centering
\includegraphics[width=0.8\linewidth]{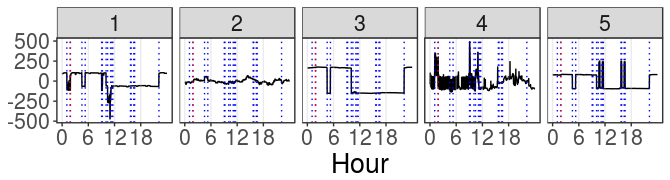}
\caption{Power: Latent source signals (1-5) recovered (black), and additive outliers (red) and level shifts (blue) detected.}
\label{fig:elec_Snew}
\end{figure*}

For comparison, we include again the recovered latent source signals with the estimated change points overlaid in Figure \ref{fig:elec_Snew}.

\end{document}